\documentclass[9pt,twocolumn,twoside]{pnas-new}

\templatetype{pnasresearcharticle}

\usepackage{pdfpages}
\usepackage{siunitx}
\usepackage{siunitx}
\usepackage{amsmath}
\usepackage{textcomp}
\usepackage{color}
\usepackage[table,dvipsnames]{xcolor}
\usepackage{soul}

\title{Harnessing Oscillatory Dynamics for Reprogrammable Mechanical Functionality}

\author[a]{Sophie Monnery}
\author[b]{Giada Risso}
\author[b]{Loucas Plado Costante}
\author[c,a,$\dagger$]{Arnaud Lazarus}
\author[b,$\dagger$]{Katia Bertoldi}

\affil[a]{Institut Jean le Rond d'Alembert, CNRS UMR7190,
Sorbonne Université, 75005 Paris, France}
\affil[b]{J.A. Paulson School of Engineering and Applied Sciences 
Harvard University, Cambridge, MA 02138, USA}
\affil[c]{Massachusetts Institute of Technology, Department of Mathematics, Cambridge, MA 02139, USA}
\affil[$\dagger$]{Corresponding authors: arnaud.lazarus@upmc.fr, bertoldi@seas.harvard.edu}

\leadauthor{Monnery}

\begin{abstract}
Structures are usually designed to perform fixed functions determined by their geometry and materials, which remain unchanged even as environmental conditions or performance requirements shift. To overcome this rigidity, researchers have introduced responsiveness directly into structures, for example through smart materials that alter their properties under external stimuli~\cite{Kim2018Printing,Wu2021Stretchable,Gladman2016Biomimetic} or jamming-based mechanisms~\cite{Brown2010UniversalGripper,Wang2019ArchitectedLattices,BrigidoGonzalez2019SwitchableStiffness}. Large deformations and instabilities have also been harnessed to realize tunable or reconfigurable responses~\cite{Wang2014HarnessingBuckling, Coulais_PhysRevLett,Reis2015Buckliphilia,Coulais2016,Arrieta2020}. Yet these approaches fall short of true mechanical reprogrammability, where structural functions can be dynamically defined, modified, and accessed on demand, akin to rewriting data on a hard drive.
To address this challenge, recent efforts have focused on building blocks that act as mechanical analogs of digital bits~\cite{yasuda2021mechanical}. Bistable units are particularly attractive because their two stable states naturally replicate binary logic~\cite{pnas_origami}. Programming arrays of such units has been pursued through two main strategies: individually addressing each bit~\cite{Chen2021ReprogrammableMetamaterial}, or exploiting interactions between neighboring units to reach desired configurations under global inputs~\cite{gorissen2017elastic,Kwakernaak2023CountingSequential,Martin_pnas}. The first offers full control but requires complex actuation schemes, while the second reduces actuation demands but often involves intricate loading sequences and nontrivial coupling design. To overcome these limitations, more recently, a dynamic control approach has been introduced that enables arbitrary transitions through global rotational driving cycles~\cite{gutierrezprieto2025dynamicdrivingenablesindependent}.
Here we introduce an alternative approach for arbitrarily defining and modifying the state of arrays of mechanical bits, inspired by recent advances in oscillating–diverging systems~\cite{lazarus2019discrete,grandi2021enhancing,grandi2023new}. We investigate arrays of pendula whose boundary conditions break symmetry, effectively transforming them into mechanical bits. When the actuation window is short relative to the natural oscillation timescales, the  state of each pendulum can be programmed solely by tuning the timing of global boundary conditions. We show that this approach enables rapid reprogramming, arbitrary information writing, and even the construction of a “mechanical piano” capable of producing user-defined note and chord sequences within just a few oscillation periods. Because it integrates seamlessly with diverse functionalities, this strategy establishes a scalable framework for dynamic, efficient, and reprogrammable mechanical systems.
\end{abstract}

\begin{document}

\maketitle
\thispagestyle{firststyle}
\ifthenelse{\boolean{shortarticle}}{\ifthenelse{\boolean{singlecolumn}}{\abscontentformatted}{\abscontent}}{}

\begin{figure}[hbt!]
    \includegraphics[width=\linewidth]{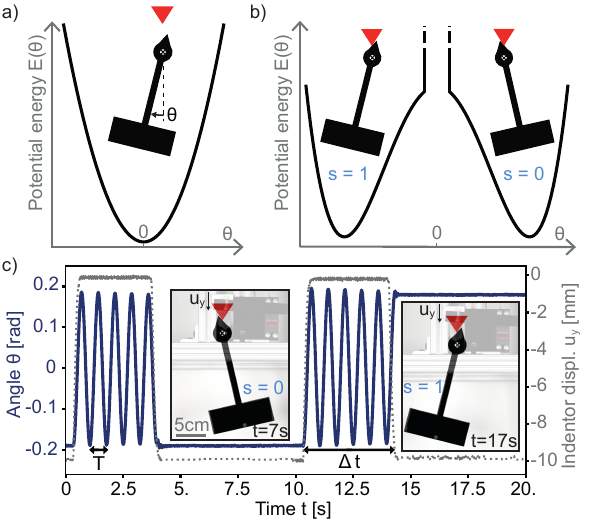}
    \caption{\textbf{Mechanical bit.}  \textbf{a)} Schematic of the globally convex energy landscape of a freely oscillating pendulum.
    \textbf{b)} Schematic of the energy landscape when the indenter is lowered, breaking the  symmetry of the system.
    \textbf{c)} Experimental time evolution of the pendulum angular motion (left axis) together with the indenter displacement $u_{y}$ (right axis). The pendulum has a natural period $T = 0.72$ s. Insets show snapshots of the pendulum locked in state  $s = 0$ at $t=7$ s and in state $s = 1$ at $t=17$ s.}
    \label{fig:one}
\end{figure}

Mechanical bits are typically non-volatile, implemented using bistable structures with a double-well energy landscape in which each stable configuration represents a digital state, either state 0 or state 1 \cite{Chen2021ReprogrammableMetamaterial,pnas_origami}. A key advantage of this approach is that the states are inherently robust and require no external energy to be maintained. However, switching between states requires an external energy input to drive the transition.
In contrast, monostable structures exhibit a globally convex energy landscape with a single stable equilibrium configuration (Fig.~\ref{fig:one}a). While they are not inherently suited for storing states, they can still be used for mechanical computing by harnessing their predictable harmonic oscillations and associating logical states with the direction of motion~\cite{yasuda2021mechanical}. However, due to damping, these oscillations decay over time and persist only when energy is continuously supplied through external excitation, thereby mimicking the behavior of volatile memory \cite{mahboob2008bit}. A promising pathway towards mechanical reprogrammable systems is to develop a hybrid system that can switch between monostable and bistable energy landscapes through a single control parameter. In this framework, the bistable (double-well) phase serves as a non-volatile regime that stores mechanical information without continuous energy input, while the monostable (single-well) phase introduces volatility and allows the system to be reprogrammed. 

Here, we present a proof-of-concept for a hybrid mechanical bit that operates by kinematically breaking the symmetry of an oscillating pendulum. Unlike conventional bistable structures, where the geometry itself gives rise to two energy minima \cite{Bazant2010Stability}, a vertically moving indenter introduces the double-well energy landscape in our system. When the indenter is raised, the pendulum oscillates freely and exhibits a single energy minimum when aligned vertically (i.e. at $\theta=0$  - Fig.~\ref{fig:one}a). Lowering the indenter reshapes the energy landscape into two potential wells separated by an effectively infinite energy barrier, thereby locking the pendulum into one of the wells, each corresponding to a distinct state (Fig.~\ref{fig:one}b). The well in which the pendulum settles is determined by its angular displacement $\theta$ at the moment of actuation.  If the indenter is lowered while $\theta$ is positive, the pendulum is captured to the left of the vertical axis, which we define as state $s = 1$ (Fig.~\ref{fig:one}c). If instead it is lowered when $\theta$ is negative, the pendulum settles to the right, corresponding to state $s = 0$ (Fig.~\ref{fig:one}c).
Thus, by precisely timing the actuation, we can deterministically set the pendulum to either state 0 or 1. Moreover, by raising the indenter, the pendulum is released into free oscillation, allowing the state to be reprogrammed by lowering the indenter again after a specific time window $\Delta t$ (Fig.~\ref{fig:one}c and Supplementary video S1). The pendulum retains the same state if   $\Delta t \in \big[(n-1)T,\, (n-1)T+\tfrac{T}{4}\big] \cup \big[nT-\tfrac{T}{4},\, nT\big]$, where $T$ is the natural period of the pendulum and $n$ is a positive integer. Conversely, it switches state if indentation occurs in $\Delta t \in \big[(n-1)T+\tfrac{T}{4},\, nT-\tfrac{T}{4}\big]$. 

\begin{figure*}
    \begin{center}
    \includegraphics[width=1\linewidth]{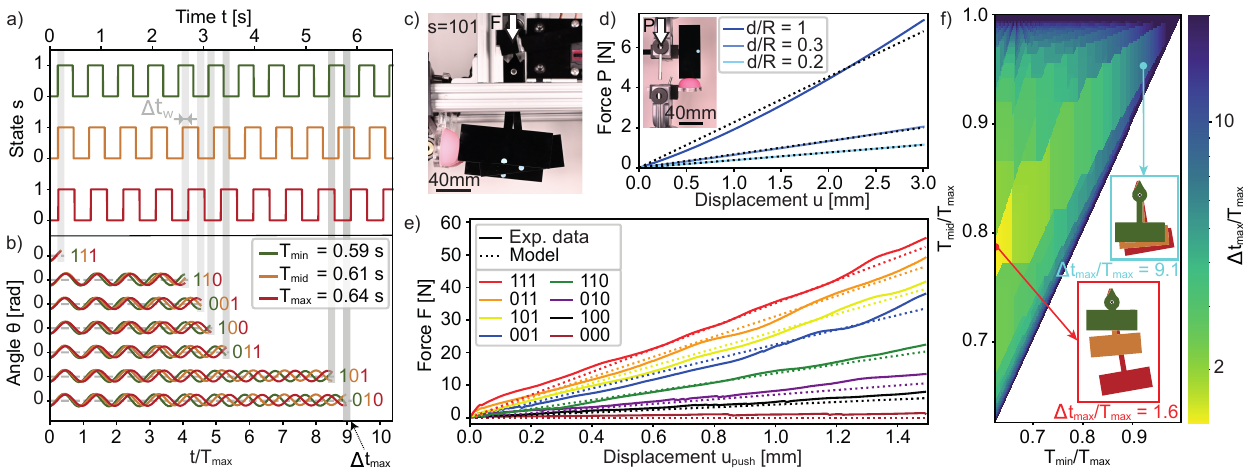}
    \caption{\textbf{State programming via  desynchronization.} 
    \textbf{a)} Temporal evolution of $s(t)$ for three pendula with $T_{\mathrm{min}} = 0.59$ s, $T_{\mathrm{mid}} = 0.61$ s, and $T_{\mathrm{max}} = 0.64$ s. Gray windows indicate intervals $\Delta t_w = 0.1$ s during which all three pendula occupy the desired state.
    \textbf{b)} Experimental temporal evolution of the  angular positions $\theta(t)$ of the three pendula.
    \textbf{c)} Experimental setup of the reprogrammable linear spring. The system is shown in state 101 with three elastomeric spheres positioned on the left side of each pendulum.
    \textbf{d)} Experimental force–displacement curves (continuous lines) for three hemispherical shells with thickness-to-radius ratio $d/R = 1$, 0.3, and 0.2 under pendulum indentation. Curves are obtained by taking the mean of three tests. Dashed lines indicate linear fits.
    \textbf{e)} Experimental force–displacement curves obtained by further lowering the indenter with the system programmed in all 8 states (continuous lines). Predictions from the summed responses of the indented shells are shown as dotted lines.
    \textbf{f)} Numerical prediction of $\Delta t_{\max} = \max(\Delta t)$ as a function of the pendula natural periods for $\Delta t_w = 0.1$ s. Blue and red markers indicate the set used in a) and the optimal set that minimizes $\Delta t _{max}$, respectively.}
    \label{fig:two}
    \end{center}
\end{figure*}

Next, we investigate the collective behavior of $N$ independent hybrid mechanical bits subjected to simultaneous actuation by a vertically moving indenter. By assigning each pendulum a distinct natural period, we leverage their gradual desynchronization to access all $2^N$ possible system states. As an illustrative example, in Fig.~\ref{fig:two} we show results for an array of three pendula with periods $T_{\mathrm{min}} = 0.59$ s, $T_{\mathrm{mid}}=0.61$ s, and $T_{\mathrm{max}}=0.64$ s. At time $t = 0$, we initialize all pendula in state $s = 0$, placing the system in state $s_{\mathrm{min}}s_{\mathrm{mid}}s_{\mathrm{max}}=000$. The pendula are then released to oscillate freely. To program a specific target state, we identify a time window $\Delta t_{\mathrm{w}}$ during which all three pendula simultaneously occupy the desired state. 
We note that the duration of the actuation window $\Delta t_{\mathrm{w}}$ must be shorter than half the smallest period in the array ($T_{\mathrm{min}}$), but large enough to ensure that the indenter can successfully lock each pendulum in place during it.  In Fig.~\ref{fig:two}a, the shaded regions mark the time intervals where the indenter must be lowered to reach all seven states from the initial 000 state when choosing $\Delta t_{\mathrm{w}} = 0.1$ s. Fig.~\ref{fig:two}b presents the measured angular trajectories (Supplementary video S2), confirming that each pendulum can indeed be trapped in the desired state. For the chosen pendula and $\Delta t_{\mathrm{w}} = 0.1$ s, state 010 requires the longest time to be accessed, with $\Delta t=\Delta t_{\mathrm{max}} = 5.85$ s $\simeq 9.1T_{\mathrm{max}}$.

The ability to efficiently program the system state enables reconfiguration of its mechanical response, which can be exploited for both functionality and state readout. As an example, we construct a reprogrammable linear spring by attaching elastomeric shells to the left side of each pendulum, such that indentation occurs only when the pendulum is in state $s=1$. After programming a state, we apply an additional displacement $u_{\mathrm{push}}$ by further lowering the indenter, causing the pendula in state $s=1$ to indent their shells while we measure the resulting force–displacement response (see Supporting Information, Section II.B).
The shells are fabricated by molding a VPS elastomer (Zhermack Elite Double 32) (see Supporting Information, Section I.B) and their mechanical response is tuned through the thickness-to-radius ratio $d/R$ (Fig.~\ref{fig:two}d and Fig.~S5). To ensure distinct responses, shells with $d/R = 0.2$, 0.3, and 1.0 (all with outer radius $R=20$ mm) are placed on the left side of the pendula with periods $T_{\mathrm{min}}$, $T_{\mathrm{mid}}$, and $T_{\mathrm{max}}$, respectively. Since each shell produces a different indentation response, each programmed state yields a distinct force–displacement curve (Fig.~\ref{fig:two}e).
This leads to two key outcomes. First, the system acts as a reprogrammable linear spring: selecting a particular bit configuration gives rise to a predictable, distinct mechanical response. Second, the one-to-one mapping between state and force–displacement curve provides a robust state readout mechanism. Because these curves can be accurately reconstructed by summing the responses of the indented shells (dotted lines in Fig.~\ref{fig:two}e), the underlying bit configuration can be identified by matching the measured response to this superposition (see Supporting Information, Section II.B), without requiring prior knowledge of the initial state or timing.

While the three pendula used in Figs.~\ref{fig:two}a–e require up to 5.85 s to reach a system state starting from 000, this time can be substantially reduced by carefully selecting their natural periods. The maximum time interval $\Delta t$ needed to access one of the seven system states from 000, denoted $\Delta t_{\mathrm{max}}$, cannot be shorter than $2^3 \Delta t_{\mathrm{w}}$. This lower bound is intuitively achieved by choosing $T_{\mathrm{min}} = 4\Delta t_{\mathrm{w}}$, $T_{\mathrm{mid}} = 8\Delta t_{\mathrm{w}}$, and $T_{\mathrm{max}} = 16\Delta t_{\mathrm{w}}$ 
(see Supporting Information, Section III.B, Fig.~S12 for details). 
However, due to fabrication tolerances and spatial constraints, the pendulum periods in our experimental setup are restricted to the range $T \in [0.5, 0.8]$ s, making it impossible to realize the theoretical optimum. Instead, we systematically explore the allowable design space to identify the set of periods that minimizes $\Delta t_{\mathrm{max}}$ within these constraints. As shown by a red dot in Fig.~\ref{fig:two}f, for $\Delta t_{\mathrm{w}} = 0.1$ s, the optimal set within the permitted range corresponds to periods $T_{\mathrm{min}}=0.63T_{\mathrm{max}}=0.5$ s and $T_{\mathrm{mid}}=0.78T_{\mathrm{max}}=0.62$ s with the longest period set to $T_{\mathrm{max}} = 0.79$ s. With those three natural periods, all eight configurations are reached within $\Delta t_{\mathrm{max}} = 1.225$ s $\simeq 1.6T_{\mathrm{max}}$ when starting from state 000 (we recall the absolute possible minimum is $\Delta t_{\mathrm{max}} = 2^3 \Delta t_{\mathrm{w}} = 0.8$ s for $\Delta t_{\mathrm{w}} = 0.1$ s).



\begin{figure}
    \includegraphics[width=\linewidth]{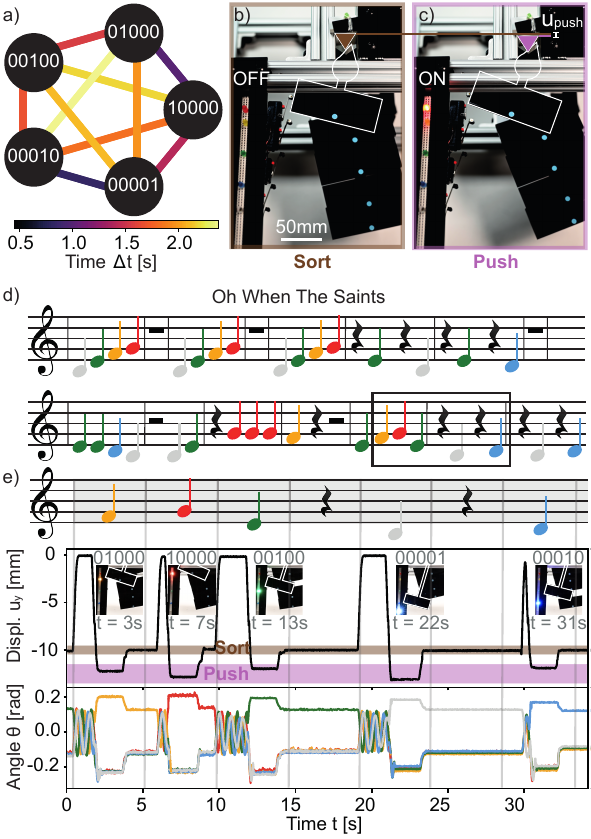}
    \caption{\textbf{Playing five-notes songs.}
\textbf{a)} Diagram of all possible transitions between the five states, with colored arrows indicating the transition times $\Delta t$.
\textbf{b)} The system sorted into state 10000 by lowering the indenter.
\textbf{c)} A further 2 mm downward displacement of the indenter brings the pendulum in state $s=1$ into contact with its switch, switching on the red LED.
\textbf{d)} Chorus of "Oh When the Saints" (34-notes sequence).
\textbf{e)} Experimental demonstration of a five-note sequence from "Oh When the Saints". Top: notes and rest times; middle: indenter displacement over time with snapshots  showing the   pendula activating the corresponding LEDs; bottom: pendulum angles $\theta$ over time.  }
    \label{fig:piano}
\end{figure}

The platform can be readily generalized to an arbitrary number of pendula and to transitions between any pair of system states. As a demonstration, we consider a system of five pendula, select five out of the $2^5$ possible states, and search for the set of pendulum periods that minimizes the maximum transition time between any pair of the selected states. The chosen states are those in which only one pendulum is in state 1 and the others are in state 0 (namely, $10000$, $01000$, $00100$, $00010$, and $00001$), yielding $N_s(N_s-1)/2 = 10$ unique state transitions, where $N_s$ is the number of selected states (Fig.~\ref{fig:piano}a). Using the minimum achievable time window of our experimental setup, $\Delta t_\mathrm{w} = 0.02$ s, we find the optimal set of pendulum periods to be $T=$ 0.50, 0.54, 0.61, 0.67, 0.76 s. As shown in Fig.~\ref{fig:piano}a, for this configuration the shortest transition occurs between states $00010$ and $00001$ with $\Delta t = 0.44$ s, while the longest occurs between states $01000$ and $00010$ with $\Delta t = 2.375$ s.

The five pendula can be viewed as the fingers of a robotic hand, each independently programmable to interact with external objects such as a keyboard or piano. To functionalize the system, we place a limit switch on the left side of each pendulum and connect it to an LED. After the pendula are programmed into a desired state (Fig.~\ref{fig:piano}b), an additional downward displacement $u_\mathrm{push}$ of about 2 mm is applied to the indenter, bringing any pendula that are in state $s=1$ into contact with their switches and thereby activating the corresponding LEDs (Fig.~\ref{fig:piano}c). To reprogram the system while maintaining consistent oscillation amplitude, the indenter is displaced upward by about 2 mm before being released into free oscillation. This procedure ensures stability and repeatability across programming cycles.

By assigning a musical note to each sensor \cite{Ben2023}, the system can be programmed to play five-notes melodies, such as "Oh When the Saints" (Fig.~\ref{fig:piano}d), with the pendula acting as actuated ‘fingers’. In this demonstration, we map the notes Do, Re, Mi, Fa, and Sol to the states $00001$, $00010$, $00100$, $01000$, and $10000$, respectively, and show that the platform can reproduce arbitrary songs composed of these five notes (Fig.~\ref{fig:piano}e, Fig.~S6 and ~S7,  Supplementary video S3 and S4). The rhythm of the song is faithfully reproduced, with the musical tempo set by the longest $\Delta t$ among the required state transitions, together with the motor control time (0.3 s) and the note-playing time (2 s). Here, $\Delta t = 2.375$ s $\simeq 3.1T_{\mathrm{max}}$  when playing Re followed by Fa, or vice versa. These factors yield a tempo of 4.7 s, as shown by the gray vertical lines separating consecutive notes in Fig.~\ref{fig:piano}d–e.

\begin{figure}
    \includegraphics[width=\linewidth]{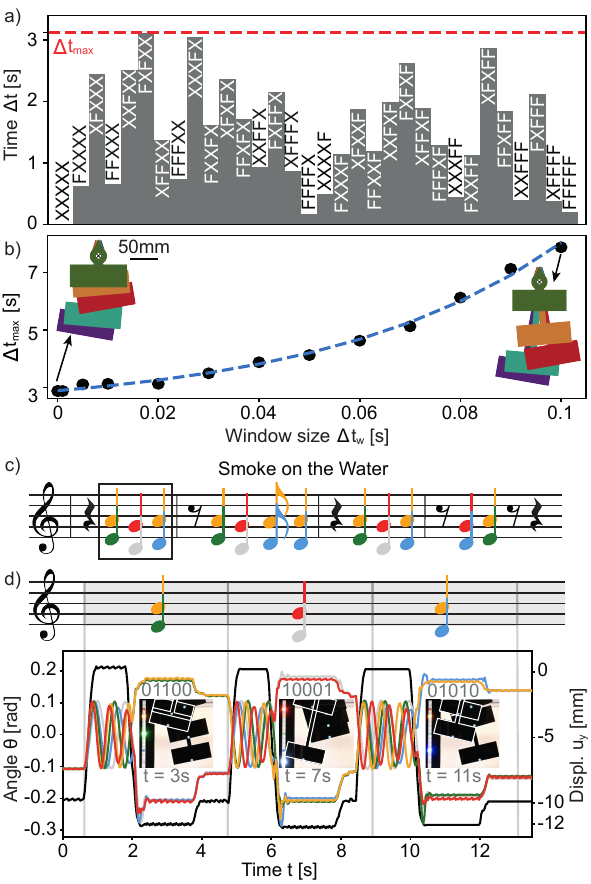}
    \caption{\textbf{Accessing all 32 states.}
\textbf{a)} Transition times $\Delta t$ for the 32 distinct state changes, using pendulum periods chosen to minimize the maximum transition time $\Delta t_{\max}$.
\textbf{b)} Dependence of $\Delta t_{\max}$ on the observation window $\Delta t_{\mathrm{w}}$.
\textbf{c)} Chorus of "Smoke on the Water" (12-chords sequence).
\textbf{d)} Experimental demonstration of a three-chords sequence from "Smoke on the Water". Right axis: indenter displacement over time (black lines). Left axis: pendulum angles $\theta$ over time (colored lines). Insets: snapshots from Supplementary Video S5.}
    \label{fig:chords}
\end{figure}

The system is not limited to the five states shown in Fig.~\ref{fig:piano}; it can be programmed to access all $2^5$ states, corresponding to 496 possible transitions. These transitions, however, are not unique, since  $\Delta t$ depends only on which pendula flip and which remain unchanged. For example, the transition from $00100$ to $01000$ takes the same time as the transition from $01101$ to $00001$, as both involve flipping the second and third pendula while leaving the others unchanged (Fig.~\ref{fig:chords}a). We denote such transitions as $XFFXX$, where ‘F’ indicates a state flip and ‘X’ denotes no change. Consequently, although 496 transitions are possible, only 32 distinct $\Delta t$  exist. We therefore search for the pendulum periods that minimize the longest of these 32 transitions, $\Delta t_{max}$. We find that for $T=$ $0.50$, $0.53$, $0.59$, $0.65$, and $0.78$ s  all 32 transitions  occur in less than $3.225$ s $\simeq 4.1T_{\mathrm{max}}$ for $\Delta t_{\mathrm{w}} = 0.02$ s (Fig.~\ref{fig:chords}a). As shown in Fig.~\ref{fig:chords}b, $\Delta t_{\max}$ depends strongly on the observation window $\Delta t_{\mathrm{w}}$: larger $\Delta t_{\mathrm{w}}$ yields systematically longer $\Delta t_{\max}$, with the set of optimal natural periods also shifting. In the limit $\Delta t_{\mathrm{w}} \to 0$, $\Delta t_{\max}$ approaches a plateau of 2.875 s.



Accessing any state within a bounded actuation time enables the system to play arbitrary sequences of notes and chords. To demonstrate this, we fabricated the set of pendula that minimizes $\Delta t_{\max}$ across all 32 possible transitions and programmed the system to reproduce a 12-chords progression from the chorus of "Smoke on the Water" (Fig.~\ref{fig:chords}c). In this demonstration, the notes Do, Re, Mi, Sol, and La are mapped to the states $00001$, $00010$, $00100$, $10000$, and $01000$, respectively, with chords formed by simultaneously activating multiple notes. For instance, the first three chords of the progression are realized by programming the system into states 01100, 10001 and 01010 (Fig.~\ref{fig:chords}d, Fig.~S8, Supplementary video S5). The rhythm of the song is faithfully captured by adjusting the resting times between transitions in proportion to the pendula sorting time, ensuring a musical tempo of 4.2 s (the longest transition used in the song is $FFXFF$ with $\Delta t = 1.9$ s).

In summary, we have demonstrated that arrays of oscillating pendula, with symmetry broken by kinematic boundary conditions, can be programmed to function as mechanical bits. The system state is dictated solely by the timing of a global actuation signal, which can be tuned to access desired configurations. This platform enables reprogrammable mechanical responses, arbitrary information encoding, and the implementation of a mechanical piano capable of executing user-defined note sequences within only a few oscillation periods of the slowest pendulum. 

By integrating multiple functionalities within a single framework, our approach establishes a new class of scalable systems that exploit dynamics for simple and efficient mechanical reprogramming. Although our experiments employed pendula as mechanical bits, the concept readily generalizes to other oscillatory systems. In particular, elastic oscillators like membranes or beams \cite{blevins2015formulas}, which are now relatively easy to make at the microscopic scale \cite{pillai2020piezoelectric},  could provide faster and more scalable reprogrammability. More broadly, the same principles may inspire analogous platforms in fluids \cite{RevModPhys1993}, chemical oscillators \cite{epstein1998introduction}, and electronic circuits \cite{strogatz1994nonlinear}, all of which inherently exhibit oscillations and multistability, underscoring the universality of dynamics as a pathway to reprogrammable matter.
 

\section*{Acknowledgments}\noindent
 K.B. acknowledges support from the Simons Collaboration on Extreme Wave Phenomena Based on Symmetries. G.R. acknowledges support from the Swiss National Science Foundation under Grant No. P500PT-217901. A.L acknowledges support from the French National Center for Scientific Research for hosting him in a CNRS delegation.
\newpage
\bibliography{References}

\clearpage
\onecolumn

\section*{\huge Supporting Information}

\setcounter{figure}{0}
\setcounter{page}{1}
\renewcommand\thefigure{S\arabic{figure}} 
\renewcommand\thepage{S\arabic{page}}
\renewcommand\theequation{S\arabic{equation}}

\renewcommand{\thesubsubsection}{\thesubsection.\arabic{subsubsection}}



\section{\Large Fabrication}\label{sec:SIfabrication}
\subsection{Fabrication  of the pendula}

\begin{figure*}[!h]
\begin{center}
\includegraphics[width = 1.0\columnwidth]{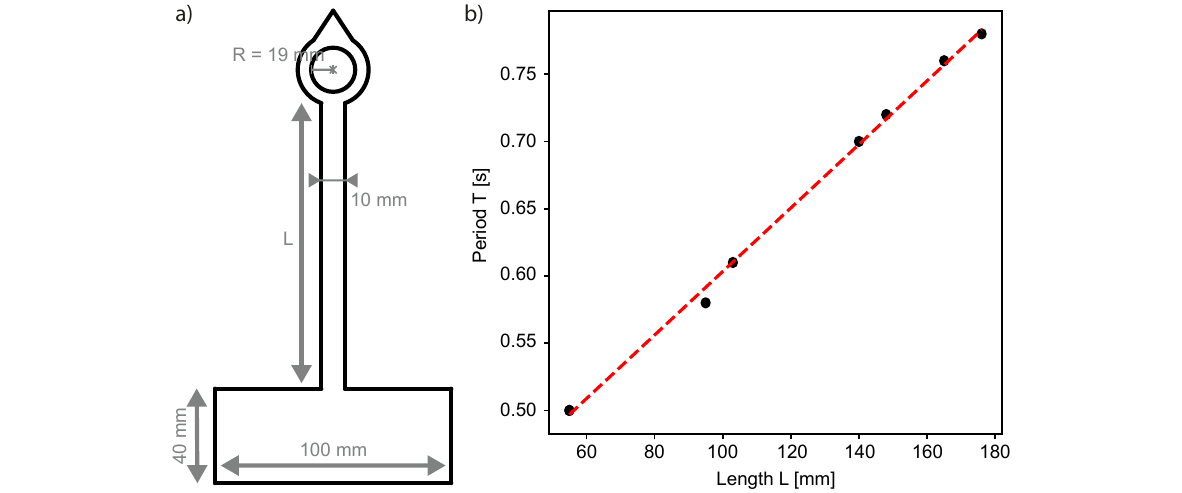} 
\caption{\label{fig:pendulum} \textbf{Design of the pendula.} \textbf{a)} Schematic of a pendulum used for laser cutting. \textbf{b)} Experimentally measured periods $T$ for pendula of varying lengths $L$ (circular markers). The dashed line shows the linear fit described by Eq.~(\ref{eq1}).}
\vspace{-15pt}
    \end{center}
\end{figure*}

The pendula are fabricated from 4 mm thick PMMA sheets, laser-cut according to the design in Fig.~\ref{fig:pendulum}a. A ball bearing is inserted into the top hole after cutting to reduce friction with the metallic support rod. The natural period is tuned by adjusting the pendulum length $L$. Figure~\ref{fig:pendulum}b reports the experimentally measured periods for seven pendula with $L \in [55,\,180]$ mm, showing a linear dependence:
\begin{equation}\label{eq1}
T = 2.4 \times 10^{-3} L + 3.7 \times 10^{-1},
\end{equation}
with $L$ in millimeters and $T$ in seconds. In this study, Eq.~(\ref{eq1}) is used to determine $L$ for a desired target period $T$.

\subsection{Fabrication of the hemispherical shells}
\begin{figure*}[!b]
\begin{center}
\includegraphics[width = 0.99\columnwidth]{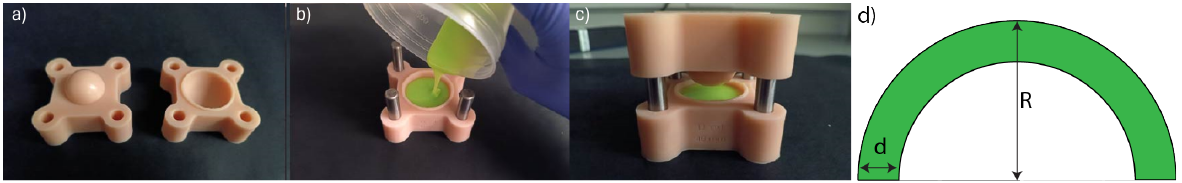} 
\caption{\label{fig:1mold} \textbf{Shell fabrication.} \textbf{a)} Photograph of the mold used in the process. \textbf{b)} The female mold part filled with uncured polyvinylsiloxane (PVS) elastomer mixture. \textbf{c)} The mold closed with pressure clamps to ensure proper alignment. \textbf{d)} Schematic of the resulting shell with thickness $d$ and radius $R$. }
\vspace{-15pt}
    \end{center}
\end{figure*}

The hemispherical shells used in this study are made of nearly incompressible polyvinylsiloxane (PVS) elastomers (Elite Double 8 from Zhermack with pink color). The shells are cast using a two-part mold designed using \textit{Onshape} and 3D printed with a Formlabs SLA 3D printer. The following step-by-step process is followed:
\begin{enumerate}
    \item We 3D print the male and female mold shown in Fig.~\ref{fig:1mold}a. A layer of Mann Release mold 200 is applied.
    \item We fill the female mold parts with the uncured polyvinylsiloxane (PVS) elastomer mixture (Fig.~\ref{fig:1mold}b).
    \item We close the molds with pressure clamps to ensure accurate alignment and keep it at room temperature (25 $^\circ$C) for 30 minutes (Fig.~\ref{fig:1mold}c).
    \item The shell is removed from the mold.
    \item Individual shells are bonded using Loctite Plastics to a circular support with radius $25$ mm, which is lasercut from 1/4'' acrylic panel. 
\end{enumerate}

For this study, we manufacture three hemispherical shells with different wall thicknesses. The outer radius of the shells is $R=20$ mm and the thickness-to-radius ratios are $d/R = 0.2$, 0.3, and 1.0.

\section{\Large Testing}\label{sec:SItesting}

\subsection{Programming states}\label{programstate}
\begin{figure*}[!h]
\begin{center}
\includegraphics[width = 1.0\columnwidth]{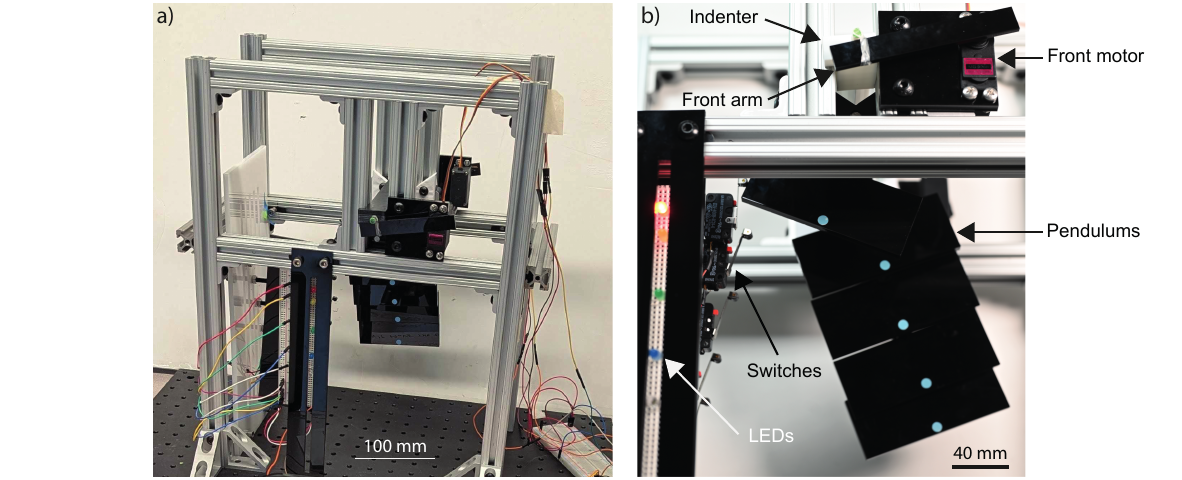} 
\caption{\label{fig:setup} \textbf{Programming states.} \textbf{a} Overview of the experimental setup. \textbf{b} Detailed view highlighting the main components.}
\vspace{-15pt}
    \end{center}
\end{figure*}

In our setup, a metallic rod with a diameter of 5 mm passes through the hole at the top of each pendulum, where ball bearings are inserted to minimize friction. As shown in Fig.~\ref{fig:setup}a, this rod is fixed to a frame built from aluminium profiles (NORCAN). A 3D-printed indenter with a triangular cross-section (22 mm wide, 15 mm high, and 250 mm long) is positioned above the pendula. Its motion is controlled by two servo motors (MG 996R) --one at the front and one at the back to prevent tilting --driven by an Arduino Mega 2560 connected to a computer. The servomotors are linked to the indenter via laser-cut arms (4 mm thick - Fig.~\ref{fig:setup}b). By setting the servo angle through an Arduino script, the indenter is translated vertically.
More specifically, the indenter displacement is controlled by the servomotor angle as follows:
\begin{itemize}
\item An angle of $1.5^\circ$ raises the indenter by approximately 1 mm.
\item An angle of $-1.5^\circ$ lowers the indenter by approximately 1 mm.
\end{itemize}


In our tests, to program a state we lower and raise the indenter by $10\,\text{mm}$, corresponding to a rotation of $\pm 15^\circ$ of the servomotor.

\subsection{Reprogrammable linear spring}
\begin{figure*}[!hpt]
\begin{center}
\includegraphics[width = 1.0\columnwidth]{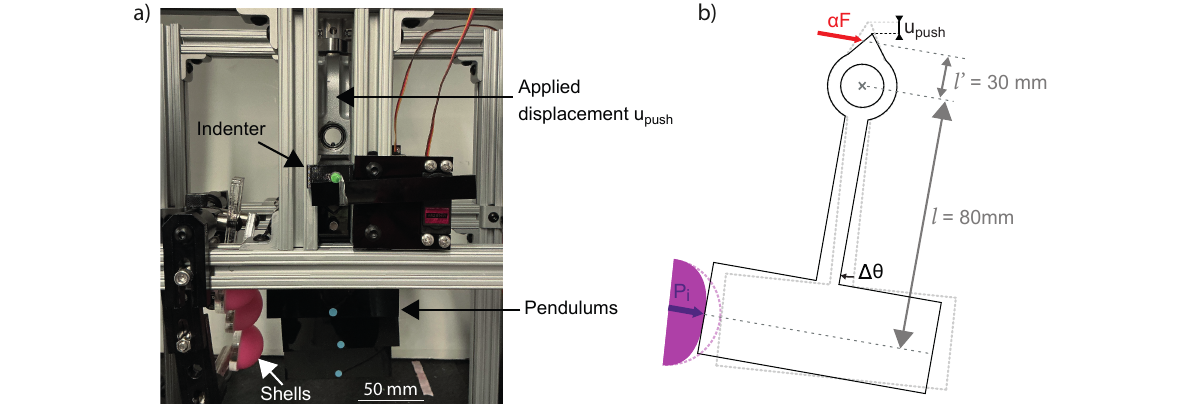} 
\caption{\label{fig:instrontesting} \textbf{Reprogrammable linear spring.} \textbf{a)} Experimental setup used to demonstrate a reprogrammable linear spring. \textbf{b)} Schematic of a pendulum indenting a shell.}
\vspace{-15pt}
   \end{center}
\end{figure*}

In Fig.~\ref{fig:instrontesting}, we show the experimental setup used to demonstrate a reprogrammable linear spring, achieved by attaching elastomeric shells to the left side of each pendulum. The response of the system is measured with an Instron uniaxial testing machine equipped with a 50~N load cell, operating under displacement control at a rate of 0.06~mm/s, chosen based on a rate-dependency study. A state is first programmed as described in Section~\ref{programstate}, after which the Instron is used to further press on the indenter while recording the reaction force.

\begin{figure*}[!hpt]
\begin{center}
\includegraphics[width = 0.99\columnwidth]{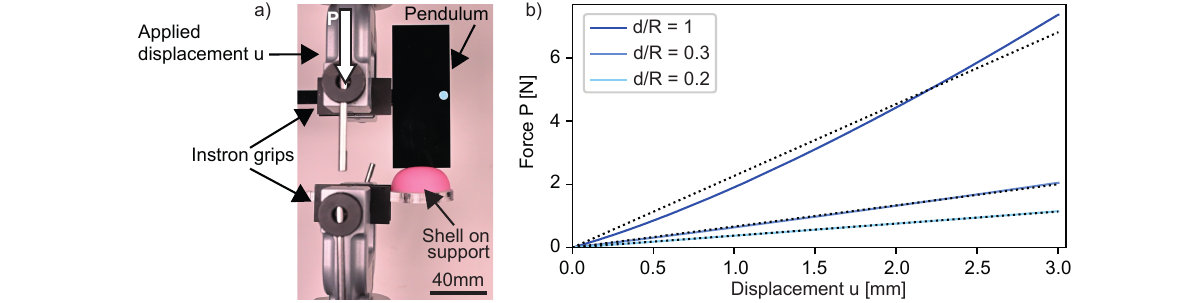} 
\caption{\label{fig:shellspoking} \textbf{Characterization of the shell response.}  
\textbf{a)} Photograph of the experimental setup used to characterize the response of an individual shell.  
\textbf{b)} Force–displacement curves of the three shells. Solid lines: experimental data; dashed lines: linear fits.}
\vspace{-15pt}
    \end{center}
\end{figure*}

To obtain eight distinct responses, three shells with different thickness-to-radius ratios were selected. The response of each shell was characterized as shown in Fig.~\ref{fig:shellspoking}, where a single pendulum was pressed against a shell while recording the applied force. The measured force–displacement curves for the three shells, displayed in Fig.~\ref{fig:shellspoking}, confirm that shells with larger $d/R$ ratios exhibit greater stiffness, as expected. The responses are well captured by linear fits (dotted lines in Fig.~\ref{fig:shellspoking}). Specifically, the response of the three considered shells is fitted as 

\begin{equation}\label{eqP}
    P_i=a_i\, u \quad \text{for $i=1,2,3$},
\end{equation}
where $P_i$ is the force in N of the $i^{th}$ shell, $u$ is the applied displacement in millimeters, and $a_i$ is a fitting parameter. We find $a_1$= $3.8e^{-1}$ N/mm for the shell with $d/R=0.2$, $a_2$=$6.7e^{-1}$ N/mm for the shell with $d/R=0.3$ and $a_3$=$2.3$ N/mm for the shell with $d/R=1.0$.

We now use these fits to predict the overall system response (Fig.~2e in the main text). 
Because the Instron applies the load to the rod connecting all pendula, the measured load $F$ is not directly equal to the sum of the shell reaction forces $P_i$ defined in Eq.~(\ref{eqP}). 
Specifically, the displacement applied by the Instron to the rod, $u_{\text{push}}$, changes the angle of each pendulum by $\Delta \theta$, which in turn generates a displacement $u$ applied by the pendulum to the shell:
\begin{equation}\label{eqU}
    u = l \sin \Delta \theta ,
\end{equation}
where $l$ is the distance between the shell--pendulum contact point and the axis of rotation of the pendulum (see Fig.~\ref{fig:instrontesting}b). 
Although each pendulum has a different length $L$, in this study the shell positions are adjusted so that $l$ is fixed at $80 \,\text{mm}$ for all pendula.  
Substituting Eq.~(\ref{eqU}) into Eq.~(\ref{eqP}) and noting that a pendulum indents a shell only when $s=1$, we obtain


\begin{equation}
P_i = 
\begin{cases}
   a_i \, l \sin \Delta \theta & \text{if } s_i = 1, \\
   0 & \text{if } s_i = 0 ,
\end{cases} 
\quad \text{for } i = 1,2,3.
\label{barPi}
\end{equation}

The rotational equilibrium of the pendula is expressed as
\begin{equation}\label{moments}  
\sum_{i=1}^3 {P}_i \, l \cos \theta = \alpha F l' ,
\end{equation}
where $l'$ is the distance between the pivot of the pendulum and the point of load application by the Instron (see Fig.~\ref{fig:instrontesting}b), and 
$\alpha F$ denotes the component of the reaction force measured by the Instron that is perpendicular to $l'$. 
Because the contact mechanics between the indenter and pendulum are complex and difficult to model explicitly, $\alpha$ is introduced as an empirical, dimensionless fitting constant.  

Next, we assume small $\Delta\theta$ and a proportionality between $u_{\text{push}}$ and $\Delta\theta$, 
\begin{equation}\label{eqAp}
    \sin\Delta\theta \approx \Delta\theta, \quad 
    \cos\Delta\theta \approx 1, \quad 
    \Delta\theta = \beta u_{\text{push}} .
\end{equation}
Substituting Eq.~(\ref{eqAp}) into Eq.~(\ref{moments}) gives
\begin{equation}
F = \sum_{i=1}^3 \frac{ \beta}{\alpha} \frac{l^2 a_i  }{l'} u_{\text{push}}.
\label{moments_final}
\end{equation}
By fitting $\beta/\alpha$ to the experimental data, we obtain $\beta/\alpha = 0.05$, which provide an accurate prediction of the global system behavior, as shown in Fig.~2e of the main text.

\clearpage
\subsection{Playing songs}

The pendula can be viewed as the fingers of a robotic hand, each independently programmable to interact with external objects such as a keyboard or piano. To functionalize the system, we place a limit switch (OMRON D3V-6G6M-3C24-K)  on the left side of each pendulum and connect it to an LED. After the pendula are programmed into a desired state, an additional downward displacement of about 2 mm is applied to the indenter, bringing any pendula that are in state $s=1$ into contact with their switches and thereby activating the corresponding LEDs.

\begin{figure*}[!hpt]
\begin{center}
\includegraphics[width = 1.0\columnwidth]{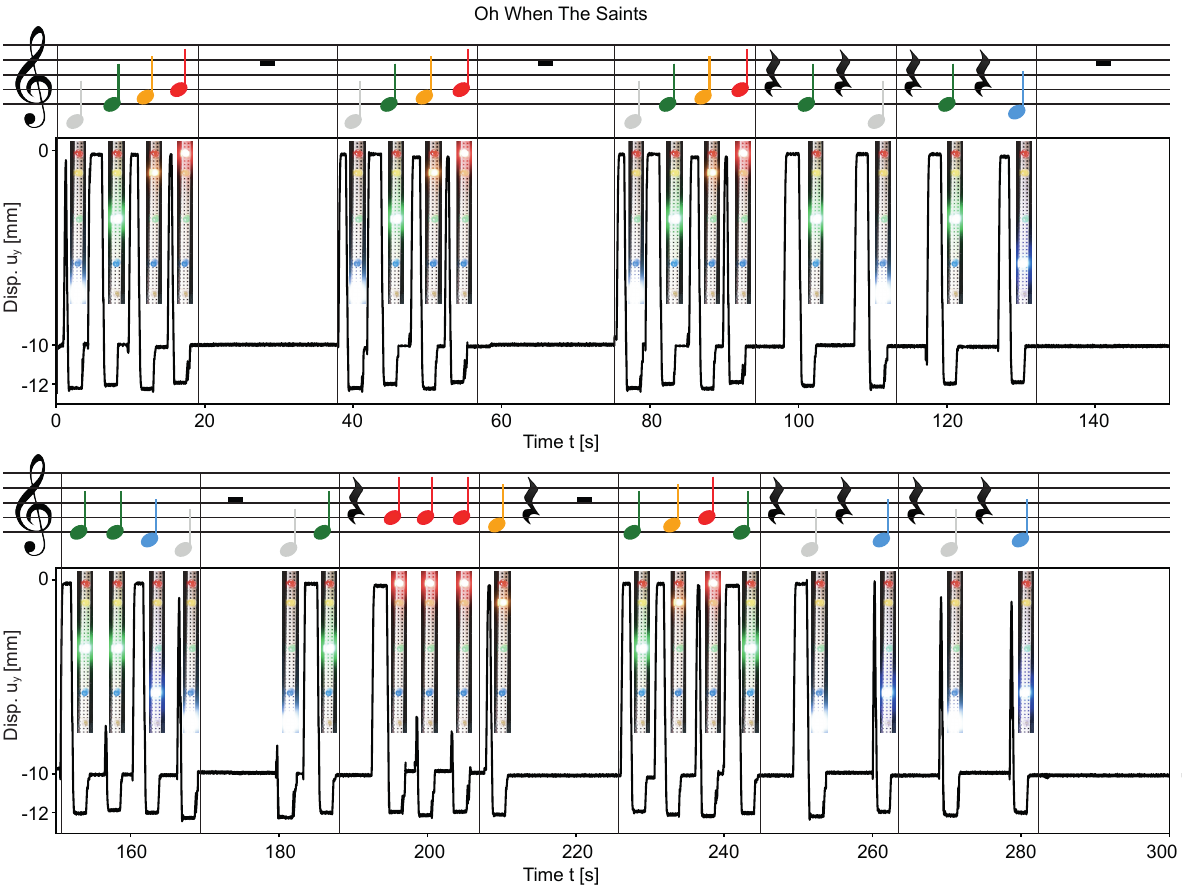} 
\caption{\label{fig:SI_OhWhen}\textbf{Playing \emph{Oh When the Saints}. } Chorus of \emph{Oh When the Saints} (34-notes sequence) together with the temporal evolution of the indenter displacement used to play it. Insets show snapshots of the LEDs during the experiment.  A recording of the experiments is provided in Supplementary Video S3.}
\vspace{-15pt}
    \end{center}
\end{figure*}

\begin{figure*}[!hpt]
\begin{center}
\includegraphics[width = 1.0\columnwidth]{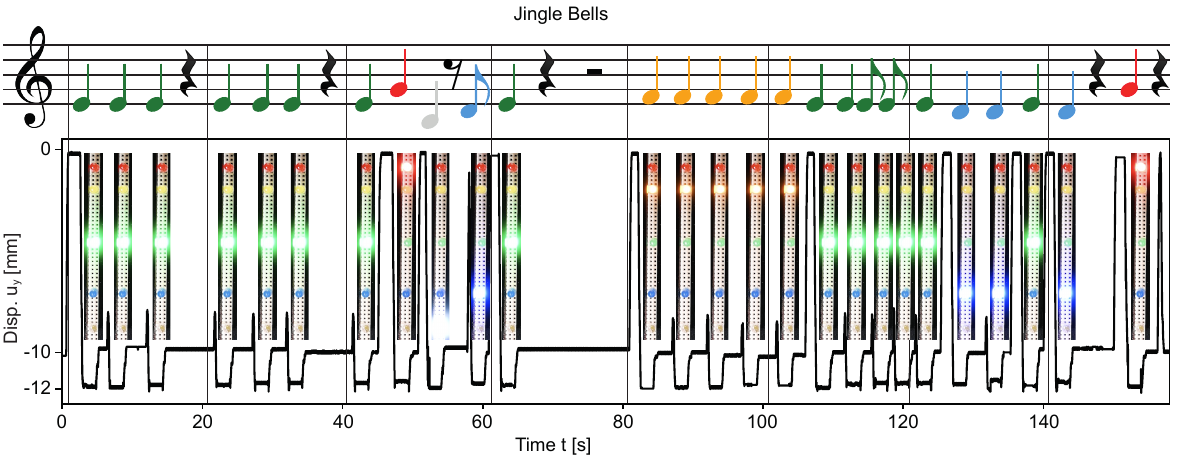} 
\caption{\label{fig:SI_JingleBells} \textbf{Playing \emph{Jingle Bells}. } Chorus of \emph{Jingle Bells} (26-notes sequence) together with the temporal evolution of the indenter displacement used to play it. Insets show snapshots of the LEDs during the experiment.  A recording of the experiments is provided in Supplementary Video S4.}
\vspace{-15pt}
    \end{center}
\end{figure*}

\begin{figure*}[!hpt]
\begin{center}
\includegraphics[width = 1.0\columnwidth]{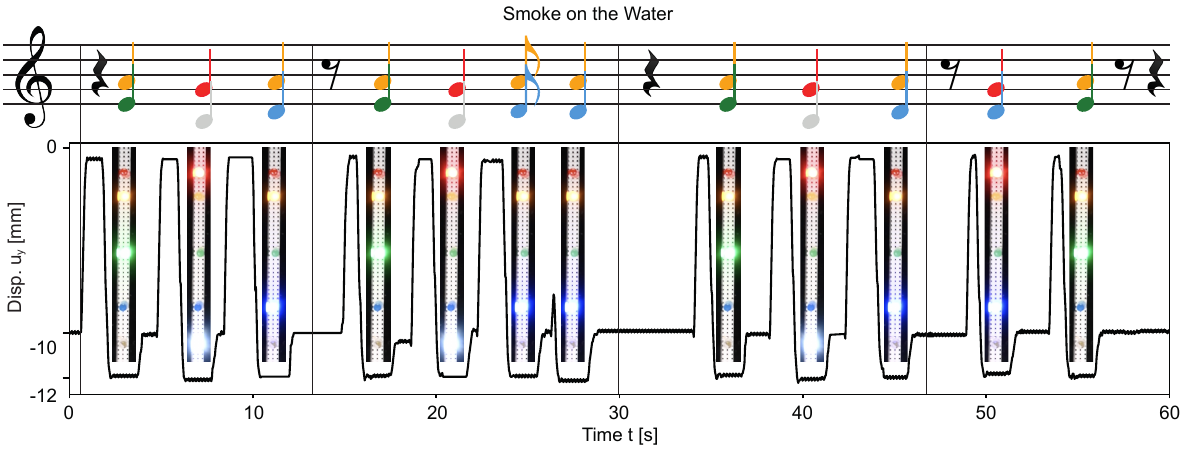} 
\caption{\label{fig:SI_Chords}  \textbf{Playing \emph{Smoke on the Water}. }  Chorus of \emph{Smoke on the Water} (12-chords sequence) together with the temporal evolution of the indenter displacement used to play it. Insets show snapshots of the LEDs during the experiment.  A recording of the experiments is provided in Supplementary Video S5.}
\vspace{-15pt}
    \end{center}
\end{figure*}

To illustrate the reprogrammability of our system, we demonstrate it by playing three songs: "Oh When the Saints" (Fig.~\ref{fig:SI_OhWhen}), "Jingle Bells"(Fig.~\ref{fig:SI_JingleBells}), and "Smoke on the Water"(Fig.~\ref{fig:SI_Chords}). The first two songs use five notes (Do, Re, Mi, Fa, and Sol), which we map to the states $00001$, $00010$, $00100$, $01000$, and $10000$, respectively. These two songs are performed using pendula with periods $T = 0.50$, $0.54$, $0.61$, $0.67$, and $0.76$ s. "Smoke on the Water" incorporates chords built from the notes Do, Re, Mi, Sol, and La, which are mapped to the states $00001$, $00010$, $00100$, $10000$, and $01000$, respectively, and is played with pendula of periods $T = 0.50$, $0.53$, $0.59$, $0.65$, and $0.78$ s.

Figures \ref{fig:SI_OhWhen}, \ref{fig:SI_JingleBells}, and \ref{fig:SI_Chords} show the complete musical notes alongside the corresponding indenter displacement $u_{y}$ and snapshots of the illuminated LEDs for each note or chord. The vertical black lines in the plots indicate the musical tempo. To faithfully reproduce the rhythm of each song, the tempo was set proportional to the longest transition time $\Delta t$ required among all state changes. For "Oh When the Saints" and "Jingle Bells", $\Delta t = 2.375$ s, and for "Smoke on the Water", $\Delta t = 1.9 $ s. The effective tempo is then calculated as the sum of three contributions: the longest state transition time ($\Delta t$), the motor control and delay time (0.3 s), and the actuation time required for pressing (2 s), resulting in 4.7 s for "Oh When the Saints" and "Jingle Bells" and 4.2 s for "Smoke on the Water". 

\clearpage
\section{\Large Numerical algorithm}\label{sec:SIalgorithm}

\subsection{Algorithm to determine $\Delta t$}

\begin{figure}[!hpt]
\begin{center}
\includegraphics[width = 0.9\columnwidth]{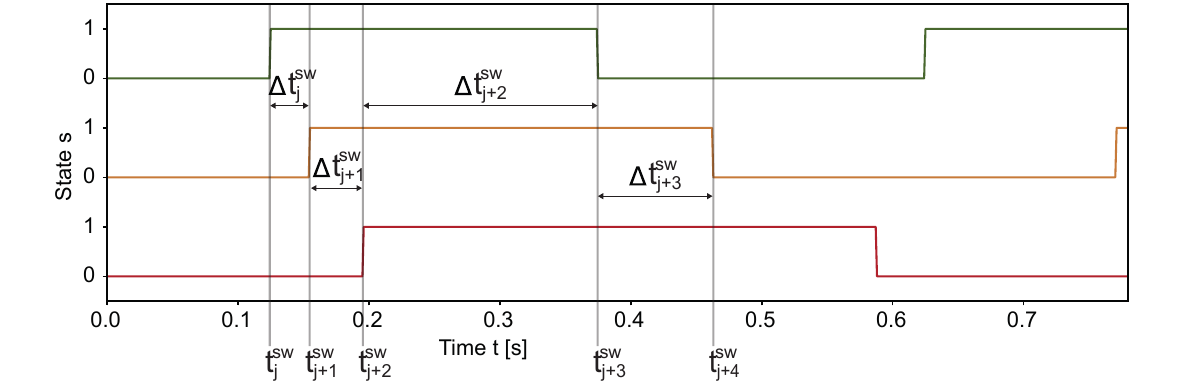} 
\caption{\label{fig:code} \textbf{Algorithm to determine $\Delta t$.} Square-wave signals $s_i(t)$, switching times $t^{\mathrm{sw}}$, and intervals $\Delta t^{\mathrm{sw}}$ for a system with $N=3$ pendula.
}
\vspace{-15pt}
    \end{center}
\end{figure}

For a system of $N$ pendula with periods $T_1, T_2, \ldots, T_N$, we determine the time interval $\Delta t$ at which the indenter should be lowered to program a desired state $s_1s_2\ldots s_N$. This is done using a Python script that proceeds as follows (see Fig.~\ref{fig:code}):  

\begin{itemize}  
  \item \textbf{Step 1:} Represent the angular motion of each pendulum as a square-wave signal $s_i(t)$, where  
  \[
    s_i(t) = \begin{cases}  
      1, & \theta_i > 0 \quad \text{(pendulum left of vertical)}, \\  
      0, & \theta_i < 0 \quad \text{(pendulum right of vertical)}.  
    \end{cases}  
  \]  

  \item \textbf{Step 2:} Compute the switching times of each signal $s_i(t)$:  
  \[
    t^{\mathrm{sw}} = \begin{cases}  
      n T_i + \tfrac{T_i}{4}, \\  
      n T_i + \tfrac{3T_i}{4},  
    \end{cases}  
    \qquad i = 1,2,\ldots,N,\;\; n=0,1,2,\ldots  
  \]  

  \item \textbf{Step 3:} Collect all switching times $t^{\mathrm{sw}}$ into a single list and sort them in ascending order.  

  \item \textbf{Step 4:} For each consecutive pair $(t^{\mathrm{sw}}_j, t^{\mathrm{sw}}_{j+1})$ compute their separation  
  \[
    \Delta t^{\mathrm{sw}}_j = t^{\mathrm{sw}}_{j+1} - t^{\mathrm{sw}}_j.  
  \]  

  \begin{itemize}  
    \item[$\diamond$] If $\Delta t^{\mathrm{sw}}_j < \Delta t_{\mathrm{w}}$, no state is considered accessible (since each state must persist for at least $\Delta t_{\mathrm{w}}$). Continue with the next pair of switching times.  

    \item[$\diamond$] If $\Delta t^{\mathrm{sw}}_j \geq \Delta t_{\mathrm{w}}$, then the system is in an accessible state during the interval $t^{\mathrm{sw}}_j < t < t^{\mathrm{sw}}_{j+1}$. This state is given by $s_1(t_j^{\mathrm{sw}})s_2(t_j^{\mathrm{sw}})\ldots s_N(t_j^{\mathrm{sw}})$.  

    \begin{itemize}  
      \item[$\star$] If this matches the desired state, store the corresponding switching time.  
      \item[$\star$] Otherwise, proceed to the next pair of switching times.  
    \end{itemize}  
  \end{itemize}  
\end{itemize}

To obtain $\Delta t$ for a list of $K$ desired states, we repeat the procedure $K$ times, once for each state. This yields a set of $\Delta t$ values, from which we determine $\Delta t_{\mathrm{max}}$ as the maximum over all $K$ results.

\subsection{Minimizing the time to program states}

\begin{figure}[!hpt]
\begin{center}
\includegraphics[width = 1\columnwidth]{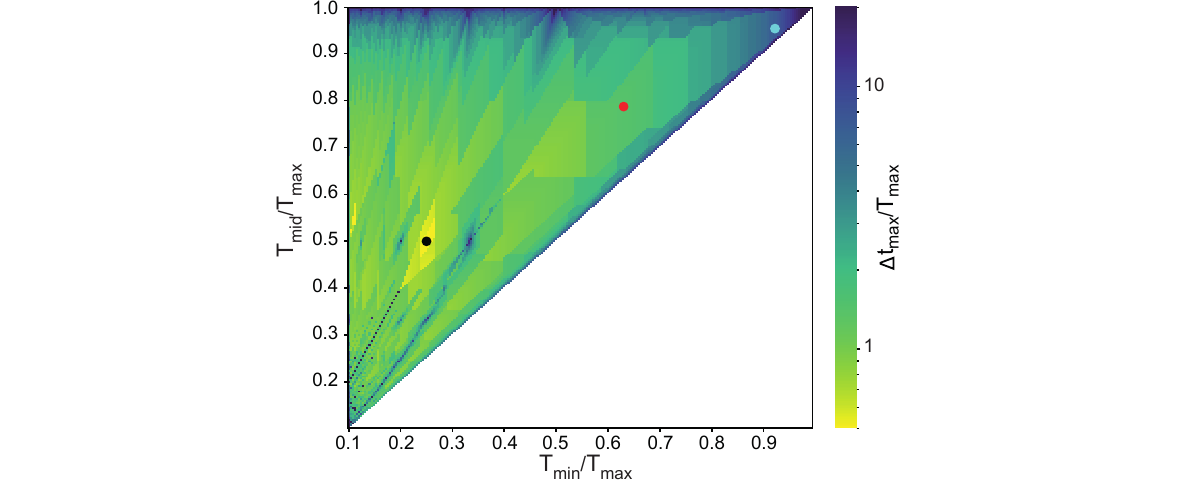} 
\caption{\label{fig:Map} \textbf{$\Delta t_{\max}$ for $N=3$ pendula as a function of $T_{min}$ and $T_{mid}$.} Numerically predicted evolution of $\Delta t_{\max}$ as a function of $T_{min}$ and $T_{mid}$ for $\Delta t_{\mathrm{w}}=0.1$ s. The blue marker in the plot corresponds to the pendula considered in Fig.~2a of the main text. The red marker indicates the optimal set of pendula that minimizes $\Delta t_{\mathrm{max}}$ within the experimental constraints (see also Fig.~\ref{fig:threeOpt}). The black marker indicates the global minimum (see also Fig.~\ref{fig:AbsoluteMinimum}).}
\vspace{-15pt}
    \end{center}
\end{figure}

\begin{figure*}[!hpt]
\begin{center}
\includegraphics[width = 1\columnwidth]{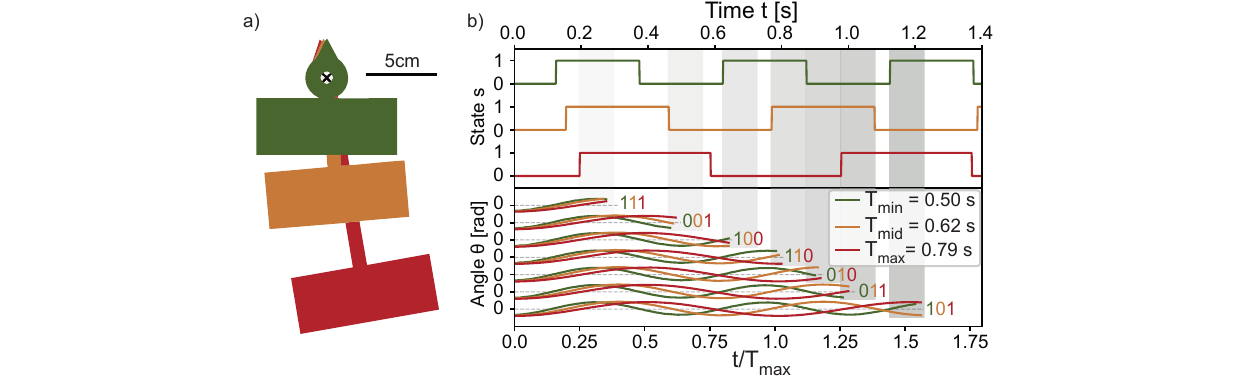} 
\caption{\label{fig:threeOpt} \textbf{Optimal set of 3 pendula experimentally accessible.} \textbf{a)} Schematics of the pendula.
\textbf{b)} Top:  Temporal evolution of $s(t)$ for three pendula with $T_{\mathrm{min}}=0.50$ s, $T_{\mathrm{mid}}=0.62$ s and $T_{\mathrm{max}}=0.79$ s. The light gray windows indicate the minimal time window $\Delta t_{\mathrm{w}}=0.1$ s in which we can observe each configuration. Bottom: Experimental temporal evolution of the angular positions $\theta(t)$ of the three
pendula.}
\vspace{-15pt}
    \end{center}
\end{figure*}

Our numerical algorithm enables a systematic investigation of how the pendulum periods affect the time required to program the system into a desired state. As an example, Fig.~\ref{fig:Map} reports numerical results for a system of three pendula initially set in state 000. We plot the time associated with the longest transition to reach  the seven accessible states from 000, $\Delta t_{\mathrm{max}}$, as a function of the pendulum periods for $\Delta t_w=0.1$ s. To generate this plot, we ran the algorithm described above 25200 times with different combinations of pendulum periods, where $T_{\mathrm{min}} \in [0.1,\,1]$ and $T_{\mathrm{mid}} \in [0.1,\,1]$. The results reveal that the pendulum periods have a strong influence on $\Delta t_{\mathrm{max}}$. The blue marker in the plot corresponds to the pendula considered in Fig.~2a of the main text (with $T_{\mathrm{min}} = 0.59$ s, $T_{\mathrm{mid}} = 0.61$ s, $T_{\mathrm{max}} = 0.64$ s), whereas the red marker indicates the optimal set of pendula that minimizes $\Delta t_{\mathrm{max}}$ within the experimental constraints ($T \in [0.5, 0.8]$ s) that correspond to periods $T_{\mathrm{min}} = 0.50$ s, $T_{\mathrm{mid}}=0.62$ s, and $T_{\mathrm{max}}=0.79$ s. Results for this optimal set of pendula are  shown in Fig.~\ref{fig:threeOpt}.

\begin{figure}[!hpt]
\begin{center}
\includegraphics[width = 1\columnwidth]{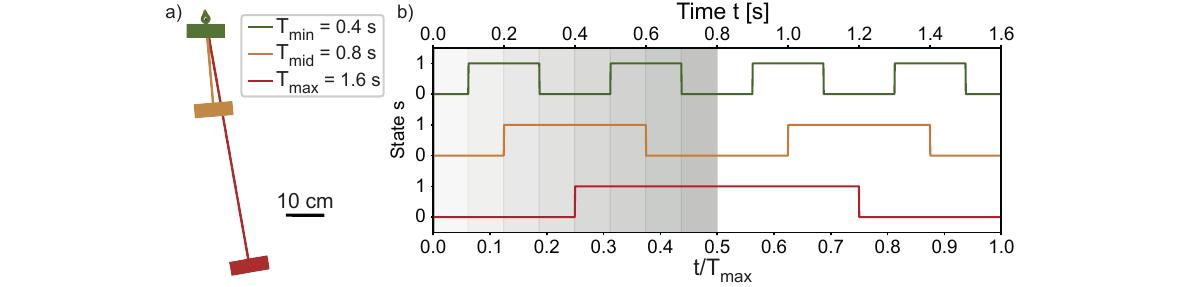} 
\caption{\label{fig:AbsoluteMinimum} \textbf{Optimal set of 3 pendula.} \textbf{a)} Schematics of the pendula.
\textbf{b)} Temporal evolution of $s(t)$ for three pendula with $T_{\mathrm{min}}=0.4$ s, $T_{\mathrm{mid}}=0.8$ s and $T_{\mathrm{max}}=1.6$ s. The light gray windows indicate the minimal time window $\Delta t_{\mathrm{w}}=0.1$ s in which we can observe each configuration.}
\vspace{-15pt}
    \end{center}
\end{figure}

Interestingly, if the experimental constraints $(T \in [0.5, 0.8]\,\text{s})$ are relaxed, the maximum interval can be further reduced to $\Delta t_{\mathrm{max}} = 2^3 \Delta t_{\mathrm{w}} = 0.8\,\text{s}$ (black marker). This value represents a global minimum: for a given $\Delta t_{\mathrm{w}}$, it is impossible to reach all seven accessible states starting from state $000$ in less time.  
A simple strategy to always ensure access to all states within $\Delta t_{\mathrm{max}} = 2^N \Delta t_{\mathrm{w}}$ is to set the largest period as $T_{\mathrm{max}} = 2\cdot2^N \Delta t_{\mathrm{w}}$, and then successively divide by two to obtain the natural periods of the remaining pendula until all $N$ values are fixed.  
Fig.~\ref{fig:AbsoluteMinimum} illustrates this for the case $N=3$, with $\Delta t_{\mathrm{w}} = 0.1\,\text{s}$, giving $\Delta t_{\mathrm{max}} = 2^3 \times 0.1 = 0.8\,\text{s}$. The three natural periods are then set as  
\begin{equation}
T_{\mathrm{max}} = 16 \Delta t_{\mathrm{w}} = 1.6\,\text{s}, \quad  
T_{\mathrm{mid}} = \frac{T_{\mathrm{max}}}{2} = 8 \Delta t_{\mathrm{w}} = 0.8\,\text{s}, \quad  
T_{\mathrm{min}} = \frac{T_{\mathrm{mid}}}{2} = 4 \Delta t_{\mathrm{w}} = 0.4\,\text{s}.
\end{equation}  
As shown in Fig.~\ref{fig:AbsoluteMinimum}b, for such a configuration of three pendula, the gray windows align perfectly one after another.

\section{\Large Supplementary Videos}

\noindent
\textbf{Movie S1.} By precisely timing the actuation, a pendulum with period $T = 0.72\,\text{s}$ can be deterministically set to state 0 or 1.\\

\noindent
\textbf{Movie S2.} Three pendula with periods $T_{\mathrm{min}} = 0.59\,\text{s}$, $T_{\mathrm{mid}} = 0.61\,\text{s}$, and $T_{\mathrm{max}} = 0.64\,\text{s}$ are programmed to reach all seven accessible states, starting from state 000. The time required to reach these seven states can be significantly reduced by instead 
choosing $T_{\mathrm{min}} = 0.50\,\text{s}$, $T_{\mathrm{mid}} = 0.62\,\text{s}$, and $T_{\mathrm{max}} = 0.79\,\text{s}$. In both cases, we used  $\Delta t_w = 0.1\,\text{s}$ to determine the actuation times.
\\

\noindent
\textbf{Movie S3.} A set of 5 pendula with periods  $T = 0.50$, $0.54$, $0.61$, $0.67$, and $0.76$ s 
is used to play "Oh When the Saints" with $\Delta t_w = 0.02$ s.\\

\noindent
\textbf{Movie S4.} A set of 5 pendula with  periods  $T = 0.50$, $0.54$, $0.61$, $0.67$, and $0.76$ s  is used to play "Jingle Bells" with $\Delta t_w = 0.02$ s.\\

\noindent
\textbf{Movie S5.} A set of 5 pendula with periods   $T = 0.50$, $0.53$, $0.59$, $0.65$, and $0.78$ s  is used to play "Smoke on the Water" with $\Delta t_w = 0.02$ s.\\


%

\end{document}